\documentclass{rmaa}

\pdfoutput=1

\usepackage{graphicx}
\usepackage{paralist}
\usepackage{psfrag,color}
\usepackage[latin1]{inputenc}

\title{Photometry of Polar-Ring Galaxies} 

\author{
  Arturo Godínez-Martínez\altaffilmark{1}, 
  Alan M. Watson\altaffilmark{1}, 
  Lynn D. Matthews\altaffilmark{1}, 
  and 
  Linda S. Sparke\altaffilmark{3}
}

\shortauthor{Godínez-Martínez et al.}
\shorttitle{Photometry of Polar Ring Galaxies}

\listofauthors{
  A. Godínez-Martínez, 
  A. M. Watson, 
  Lynn D. Matthews
  \& 
  L. S. Sparke
}

\altaffiltext{1}{Centro de Radioastronomía y Astrofísica, Universidad Nacional Autónoma de México.}
\altaffiltext{2}{Harvard-Smithsonian Center for Astrophysics}
\altaffiltext{3}{Department of Astronomy, University of Wisconsin-Madison}

\fulladdresses{
\item
A. Godínez-Martínez and A.~M.~Watson: Centro de Radioastronomía y
Astrofísica, Universidad Nacional Autónoma de México, Apartado Postal
3-72 (Xangari), 58089 Morelia, Michoacán, México
(arturogm@gmail.com, a.watson@astrosmo.unam.mx).
\item
L.~D. Matthews: Harvard-Smithsonian Center for Astrophysics, Cambridge,
MA 02138, USA (lmatthew@cfa.harvard.edu).
\item
L.~S.~Sparke: Department of Astronomy, University of Wisconsin-Madison,
475 North Charter Street, Madison, WI 53706-1582, USA
(sparke@astro.wisc.edu). }

\indexauthor{Godínez-Martínez, A.}
\indexauthor{Watson, A. M.}
\indexauthor{Matthews, L. D.}
\indexauthor{Sparke, L. S..}

\abstract{We have obtained photometry in $B$ and $R$ for seven confirmed
or probable polar-ring galaxies from the Polar-Ring Catalog of Whitmore
et al.\@ (1990). The rings show a range of colors from $B-R \approx 0.6$
to $B-R\approx 1.7$. The bluest rings have bright \ion{H}{II} regions,
which are direct evidence for recent star formation. The minimum age of
the reddest ring, that in PRC B-20, is somewhat uncertain because of a
lack of knowledge of the internal reddening and metallicity, but appears
to be at least 1.2 Gyr. As such, this ring is likely to be stable for at least
several rotation periods. This ring is an excellent candidate for future
studies that might better determine if it is truly old.}

\resumen{ Se obtuvo fotometría en $B$ y $R$ en siete galaxias con anillo
polar, probables o confirmadas, del Catálogo de Anillos Polares (PRC) de
Whitmore et al.\@ (1990). Los anillos tienen un amplio intervalo de
colores de $B-R \approx 0.6$ hasta $B-R\approx 1.7$. Los anillos más
azules tienen brillantes regiones \ion{H}{II}, las cuales son evidencia
directa de formación estelar reciente. La edad mínima del anillo más
rojo, el de PRC B-20, es algo incierta debido a la falta de conocimiento
del enrojecimiento interno y metalicidad en este sistema, pero parece
ser al menos de 1.2 Gyr. Por lo tanto, este anillo parece ser estable,
al menos por varios periodos rotacionales. Este anillo es un excelente
candidato a estudios futuros que deben determinar de mejor manera si en
realidad es un anillo viejo.}

\addkeyword{galaxies: halos}
\addkeyword{galaxies: photometry}
\addkeyword{galaxies: stellar content}

\begin{document}

\maketitle

\section{Introduction}

Polar-Ring Galaxies are systems with two kinematically distinct
components (Schechter \& Gunn 1978). The central component is an
apparently normal galaxy, usually an S0. The second is a ring of gas,
dust, and stars whose orbital plane is close to orthogonal to that of
the host galaxy. The rings are thought to be the result of an
interaction between the host galaxy and a donor (Schweizer, Whitmore, \&
Rubin 1983).

The ages of the rings are interesting from at least two points of view.
First, by combining the mean age of the rings with the current frequency
of polar-ring galaxies, once can estimate the frequency with which polar
rings are formed. By comparing this to the frequency of all types of
interactions, one might hope to learn which kinds of interactions
produce polar rings. Second, it is not clear that polar rings are
stable. Inclined orbits in a oblate potential suffer differential
precession, and so a ring formed by populating such orbits would be
smeared out on a few orbital timescales. If rings are long-lived, then
it may be that they are self-gravitating or that the halos of the host
galaxies are triaxial (see the discussion in Cox \& Sparke 1996 and
Sparke 2004).

The stars currently seen in the ring almost certainly formed from the
gas of the ring (Bournaud \& Combes 2003). Therefore, their age sets a
lower limit to the age of the ring. The standard means to measure the
age of an unresolved stellar population is to interpret broadband
photometry using stellar population synthesis methods. A truly old ring
should be red, although reddening by dust could disguise a young ring as
an old one. Furthermore, the precise mapping from unreddened color to
age depends quite sensitively on the metallicity and star-formation
history.

Many polar rings are quite blue (see \S5.2) and are not good candidates
for old rings. Our aims in this work are to attempt to identify red
rings using $B-R$ photometry. We selected $B-R$ as a good compromise
between sensitivity and color information. Because of the uncertainties
introduced by dust, metallicity, and star-formation history, we do not
pretend to be able to extract precise ages on the basis of a single
color. However, red rings are excellent candidate subjects for future
investigations to confirm whether they are indeed old rings.

\section{Observations}

We observed seven polar-ring galaxies from the Polar Ring Catalog
(Whitmore et al.\@ 1990) using the 1.5\,m telescope of the Observatorio
Astronómico Nacional on Sierra San Pedro Mártir, Baja California,
Mexico, the SITe1 $1024\times1024$ CCD (binned $2\times2$ to give a
scale of 0\farcs51/pixel and a field of view of 4\farcm3) and the $B2$
(1\,mm CG285 + 1\,mm BG18 + 2\,mm BG12) and $R2$ (2\,mm KG3 + 2\,mm
OG570) filters. Table~\ref{table:log} gives names and equatorial
coordinates of each galaxy (from NED) and the dates and exposure times
of the observations. We obtained bias and twilight flat field exposures
each night.

\begin{table*}\centering
\footnotesize
\setlength{\tabnotewidth}{\columnwidth}
\tablecols{5}
\caption{Observing Log}
\label{table:log}
\begin{tabular}{lclcc}
\toprule
Galaxy & Coordinates (J2000) & \multicolumn{1}{c}{Date} & Exposures in $B2$& Exposures in $R2$\\
\midrule
PRC A-01, A 0136-0801        &01 38 55.2 $-$07 45 56 &2001 September 19    &$2\times1000$\,s &$2 \times 500$\,s \\
PRC A-04, UGC 7576           &12 27 41.8 $+$28 41 53 &2001 April 26        &$2\times1000$\,s &$2 \times 500$\,s \\
PRC A-06, UGC 9796, II Zw 73 &15 15 56.3 $+$43 10 00 &2001 April 24 and 30 &$4\times1000$\,s &$4 \times 500$\,s \\
PRC B-10, A 0950-2234        &09 52 53.9 $-$22 48 34 &2001 April 26 and 29 &$4\times1000$\,s &$4 \times 500$\,s \\
PRC B-17, UGC 9562, II Zw 71 &14 51 14.4 $+$35 32 32 &2001 April 24        &$2\times1000$\,s &$2 \times 500$\,s \\
PRC B-20, A 2135-2132        &21 38 20.0 $-$21 19 06 &2001 September 20    &$2\times1000$\,s &$2 \times 500$\,s \\
PRC B-21, ESO 603-G21        &22 51 22.0 $-$20 14 50 &2001 September 19    &$2\times1000$\,s &$2 \times 500$\,s \\
\bottomrule
\end{tabular}
\end{table*}

We used IRAF to reduce the exposures of the galaxies. Specifically, we
used the ccdred package to construct and apply bias and flat field
corrections, the cosmicrays task to identify automatically and
interpolate automatically over cosmic ray events, and the imedit task to
interpolate manually over remaining cosmic ray events. We shifted the
exposures in each filter into a common alignment and summed them. The
first columns of Figures~\ref{figure:A-01} to \ref{figure:B-21} show the
final images of each galaxy in $B2$ and $R2$.

The galaxies were observed in the course of observing runs in the spring
and fall of 2001 on the 1.5 meter telescope. After the fall run on the
1.5 meter telescope, we had a run on the adjacent 84 centimeter
telescope with the same SITe 1 CCD and the same $B2$ and $R2$ filters.
During these three runs we observed photometric standards from Landolt
(1983). Subsequent analysis showed that nine nights on the 1.5 meter
telescope and a further two nights on 84 centimeter telescope were
photometric. During these photometric nights we observed a total of 106
(in $B$) and 107 (in $R$) photometric standards with $B-R$ colors from
$-0.4$ to $+2.2$ at airmasses from 1.1 to 2.3. We reduced the standards
in the same way as the galaxies, except that we did not interpolate
manually over cosmic ray events.

We performed aperture photometry on the standard stars using the apphot
package. We used an aperture of diameter 15\farcs2. We fitted for the
transformation from instrumental to standard magnitudes, allowing for a
zero point, a linear extinction term, and a linear color term. We fitted
for different zero points and extinction coefficients for each night but
for a single color coefficient for the whole set of nights. Using a
single color coefficient improves the precision of the fit. (This is the
reason for including data from the 84 centimeter telescope, even though
we observed no galaxies during this run.) The RMS residuals were 2.0\%
in $B$ and 1.3\% in $R$. The extinction coefficients varied between 0.17
and 0.32 for $B2$ and 0.07 and 0.17 in $R2$.

The transformations from the natural $B2$ and $R2$ magnitudes to the
standard $B$ and $R$ magnitude were $B = B2 + 0.112(B-R)$ and $R = R2 +
0.008(B-R)$. This indicates that the $B2$ and $R2$ bandpasses are redder
than the standard bandpasses, with $B2$ being redder by about 200\,\AA.
Given that the $B2$ filter is constructed according to a recipe for
photocathodes rather than CCDs, these differences are not surprising.

\section{Model Fitting}

Polar rings are often faint compared to their host galaxies. For this
reason, we modelled and subtracted the host galaxies to allow for better
photometry of the rings.

We fitted models to the galaxies using Galfit (Peng et al.\@ 2002). This
program takes an image, an uncertainty image, and a point-spread function
(PSF) image and fits a model consisting of a tilted plane for the sky
and combinations of Sérsic profiles, exponential disks, gaussians, and
PSFs for the galaxy. The model parameters are adjusted to minimize the
reduced $\chi^2_\nu$.

We began by creating a mask for each image to indicate regions that
should not be included in the fit. These regions include stars, the
brightest parts of the ring, and the region in the galaxy PRC A-01 where
the near side of the ring passes over the galaxy and causes significant
extinction. The second columns of Figures~\ref{figure:A-01} to
\ref{figure:B-21} show these masks. We created PSF images using field
stars and uncertainty images using the known gain and read noise of the CCD.

We began modelling the galaxies with a tilted plane for the sky and a
single Sérsic or exponential disk component. We added components until
the reduced $\chi^2_\nu$ of the fit no longer decreased significantly.
The components are listed in Table~\ref{table:galfit}. See Peng et al.
(2002) for definitions of the fitted components and their parameters.
The third columns of Figures~\ref{figure:A-01} to \ref{figure:B-21} show
the images after subtracting the models of the host galaxy.

Some of the components have clear physical interpretations, but others,
especially the gaussians, do not. We suspect that such ad hoc components
appear because of errors in the PSF, dust absorption close to the center
of the galaxies, and because the components in Galfit are more
appropriate for fitting less inclined galaxies. Nevertheless, we are not
especially worried by this as our aim is not to investigate the
structure of the host galaxies but simply to remove their light. Also,
despite our attempts to exclude the rings from the fits, the models
often seem to remove the rings where they are projected against the host
galaxy. For these reasons we will restrict our photometry of rings to
the parts that are not projected over bright regions the host galaxy.

\begin{table*}\centering
\setlength{\tabnotewidth}{0.7\textwidth}
\tablecols{11}
\caption{Fitted Components}
\label{table:galfit}
\footnotesize
\begin{tabular}{lllrrrrrrrr}
\toprule
Galaxy & Filter & Component & 
\multicolumn{1}{c}{$\Delta\alpha$\tabnotemark{a}}& 
\multicolumn{1}{c}{$\Delta\delta$\tabnotemark{a}}& 
\multicolumn{1}{c}{$\Delta m$\tabnotemark{b}}& 
\multicolumn{1}{c}{$r$\tabnotemark{c}}&
\multicolumn{1}{c}{$n$\tabnotemark{d}}& 
\multicolumn{1}{c}{$q$\tabnotemark{e}}& 
\multicolumn{1}{c}{PA\tabnotemark{f}}& 
\multicolumn{1}{c}{$c$\tabnotemark{g}}\\
& 
$\chi^2_\nu$\\

\midrule
PRC A-01&$B$&Sérsic  &$ 0\farcs0$&$ 0\farcs0$&0.00&$4\farcs9$  &1.20   &0.59   &$-44.3$&$-0.32$\\
    &1.5&PSF     &$-0\farcs9$&$+0\farcs9$&1.34&\nodata&\nodata&\nodata&\nodata&\nodata\\
    &   &Gaussian&$-0\farcs3$&$-0\farcs3$&3.08& $0\farcs3$  &\nodata&0.04   &$-22.8$&$-1.76$\\
\cmidrule{2-11}
    &$R$&Sérsic  &$ 0\farcs0$&$ 0\farcs0$&0.00&$4\farcs8$  &1.39   &0.55   &$-44.7$&$-0.21$\\
    &2.0&Gaussian&$-0\farcs4$&$+0\farcs5$&1.19& $1\farcs5$  &\nodata&0.30   &$+14.0$&$-0.86$\\
    &   &PSF     &$-0\farcs9$&$-0\farcs2$&4.61&\nodata&\nodata&\nodata&\nodata&\nodata\\
\midrule
PRC A-04&$B$&Sérsic  &$ 0\farcs0$&$ 0\farcs0$&0.00&$12\farcs6$  &1.16   &0.79   &$+61.3$&$-0.64$\\
    &1.3&Sérsic  &$+0\farcs2$&$-0\farcs8$&0.47&$5\farcs5$  &0.78   &0.27   &$-46.8$&$+1.05$\\
    &   &Sérsic  &$-0\farcs1$&$-0\farcs5$&0.95& $1\farcs4$  &1.66   &0.59   &$-9.6$&$-1.25$\\
\cmidrule{2-11}
    &$R$&Sérsic  &$ 0\farcs0$&$ 0\farcs0$&0.00&$10\farcs2$  &1.76   &0.89   &$-53.8$&$-0.22$\\
    &1.2&Sérsic  &$+0\farcs2$&$-0\farcs2$&0.79&$5\farcs7$  &0.57   &0.29   &$-45.7$&$-0.04$\\
    &   &Sérsic  &$+0\farcs1$&$-0\farcs1$&1.13& $0\farcs8$  &1.41   &0.08   &$-25.8$&$ 0.00$\\
    &   &PSF     &$-0\farcs3$&$+2\farcs0$&4.75&\nodata&\nodata&\nodata&\nodata&\nodata\\
\midrule
PRC A-06&$B$&Sérsic  &$ 0\farcs0$&$ 0\farcs0$&0.00&$7\farcs4$  &1.35   &0.67   &$-33.2$&$-0.12$\\
    &1.8&Sérsic  &$-2\farcs9$&$+1\farcs8$&0.83&$7\farcs3$  &0.37   &0.22   &$-55.2$&$ 0.00$\\
    &   &Sérsic  &$-1\farcs8$&$+1\farcs2$&1.40& $1\farcs3$  &0.03   &0.82   &$ +77.2$&$-0.51$\\
    &   &PSF     &$-2\farcs4$&$+1\farcs4$&2.05&\nodata&\nodata&\nodata&\nodata&\nodata\\
\cmidrule{2-11}
    &$R$&Sérsic  &$ 0\farcs0$&$ 0\farcs0$&0.00&$5\farcs6$  &0.83   &0.27   &$-54.5$&$ 0.00$\\
    &1.4&Sérsic  &$-0\farcs9$&$+0\farcs6$&0.36&$10\farcs7$  &0.24   &0.65   &$-30.4$&$-0.71$\\
    &   &Sérsic  &$+0\farcs0$&$+0\farcs4$&1.23& $1\farcs1$  &0.88   &1.00   &$-70.2$&$-0.72$\\
    &   &PSF     &$-0\farcs8$&$+0\farcs4$&1.47&\nodata&\nodata&\nodata&\nodata&\nodata\\
\midrule
PRC B-10&$B$&Sérsic  &$ 0\farcs0$&$ 0\farcs0$&0.00&$3\farcs2$   &1.79   &0.59   &$-75.7$&$+0.06$\\
    &1.1&\\
\cmidrule{2-11}
    &$R$&Sérsic  &$ 0\farcs0$&$ 0\farcs0$&0.00& $3\farcs2$  &1.18   &0.57   &$-76.4$&$+0.43$\\
    &1.0&PSF     &$-0\farcs2$&$+0\farcs0$&2.15&\nodata&\nodata&\nodata&\nodata&\nodata\\
\midrule
PRC B-17&$B$&Sérsic  &$ 0\farcs0$&$ 0\farcs0$&0.00&$13\farcs1$  &1.70   &0.46   &$+34.1$&$+0.16$\\
    &1.8&Exp.    &$+0\farcs5$&$+0\farcs9$&0.07&$10\farcs2$  &\nodata&0.50   &$-37.7$&$-0.20$\\
    &   &Sérsic  &$+0\farcs3$&$+1\farcs8$&2.62& $1\farcs4$  &1.20   &0.75   &$-79.6$&$-0.42$\\
\cmidrule{2-11}
    &$R$&Exp.    &$ 0\farcs0$&$ 0\farcs0$&0.00&$9\farcs6$  &\nodata&0.57   &$-37.7$&$-0.34$\\
    &1.2&Sérsic  &$+0\farcs3$&$-1\farcs0$&0.52&$10\farcs3$  &1.59   &0.54   &$+30.2$&$+0.60$\\
    &   &Sérsic  &$+0\farcs2$&$+0\farcs8$&3.02& $1\farcs6$  &2.07   &0.74   &$-67.7$&$-0.56$\\
\midrule
PRC B-20&$B$&Exp.    &$ 0\farcs0$&$ 0\farcs0$&0.00&$5\farcs3$  &\nodata&0.92   &$-63.6$&$-0.14$\\
    &1.2&Sérsic  &$-0\farcs6$&$+0\farcs2$&1.40& $2\farcs9$  &0.72   &0.61   &$+4.7$&$+0.70$\\
    &  &PSF      &$-0\farcs4$&$+0\farcs1$&2.68&\nodata&\nodata&\nodata&\nodata&\nodata\\
\cmidrule{2-11}
    &$R$&Exp.    &$ 0\farcs0$&$ 0\farcs0$&0.00&$5\farcs3$  &\nodata&0.91   &$-68.2$&$-0.12$\\
    &1.6&Sérsic  &$-0\farcs2$&$+0\farcs2$&0.96& $2\farcs5$  &1.34   &0.64   &$+8.8$&$+0.55$\\
    &   &PSF     &$-0\farcs2$&$+0\farcs3$&2.60&\nodata&\nodata&\nodata&\nodata&\nodata\\
\midrule
PRC B-21&$B$&Sérsic  &$ 0\farcs0$&$ 0\farcs0$&0.00&$8\farcs2$  &0.37   &0.97   &$+65.9$&$+0.77$\\
    &1.6&Sérsic  &$+1\farcs9$&$+1\farcs8$&2.36& $2\farcs8$  &0.32   &0.67   &$-63.1$&$-0.18$\\
\cmidrule{2-11}
    &$R$&Sérsic  &$ 0\farcs0$&$ 0\farcs0$&0.00&$7\farcs3$  &0.47   &0.96   &$+70.4$&$+0.89$\\
    &2.3&Sérsic  &$+0\farcs9$&$+0\farcs6$&2.39& $2\farcs1$  &0.68   &0.42   &$-64.6$&$+0.50$\\
\bottomrule

\tabnotetext{a}{Offsets between the centers of the components.}
\tabnotetext{b}{Magnitude relative to the brightest component (which has
$\Delta m \equiv 0$).}
\tabnotetext{c}{$r$ is $r_\mathrm{e}$ for exponenential profiles, $r_\mathrm{s}$ for
  Sérsic profiles, and FWHM for Gaussian profiles.}
\tabnotetext{d}{$1/n$ is the Sérsic profile exponent.}
\tabnotetext{e}{$q$ is the axis ratio.}
\tabnotetext{f}{PA is the position angle.}
\tabnotetext{g}{$c$ is the {\it diskiness} (positive) or {\it boxiness} (negative) parameter.}
\end{tabular}
\end{table*}

\section{Photometry}

\subsection{Galactic Extinction}

\begin{table}\centering
\setlength{\tabnotewidth}{\columnwidth}
\tablecols{3}
\caption{Assumed Galactic Extinction}
\label{table:extinction}
\begin{tabular}{lcc}
\toprule
Galaxy &
$A_{B2}$&
$A_{R2}$\\
\midrule
PRC A-01 &0.110&0.071 \\
PRC A-04 &0.089&0.058 \\
PRC A-06 &0.110&0.072 \\
PRC B-10 &0.192&0.124 \\
PRC B-17 &0.052&0.034 \\
PRC B-20 &0.178&0.116 \\
PRC B-21 &0.137&0.089 \\
\bottomrule
\end{tabular}
\end{table}

We correct our photometry for Galactic extinction using values from NED,
which are taken from appendix B of Schlegel et al.\@ (1998). Extinction
corrections must be calculated for the wavelengths of the natural
system. Since our $R2$ filter is very similar to the the standard $R$,
we use $A_{R2} = A_R$. However, since our $B2$ bandpass is about
200\,{\AA} redder than the standard bandpass, we use $A_{B2} = 0.8A_B +
0.2A_V$. Table~\ref{table:extinction} lists are adopted extinction
values. The largest reddening is 0.07 for PRC B-10.

\subsection{Host Galaxies}

\begin{table*}\centering
\setlength{\tabnotewidth}{0.8\textwidth}
\tablecols{10}
\caption{Photometry}
\label{table:photometry}
\begin{tabular}{lccccccccc}
\toprule
Galaxy && 
\multicolumn{2}{c}{Aperture\tabnotemark{a}}&&
\multicolumn{2}{c}{Model\tabnotemark{b}}&&
\multicolumn{2}{c}{Ring\tabnotemark{c}}\\
\cmidrule{3-4}
\cmidrule{6-7}
\cmidrule{9-10}
&&
\multicolumn{1}{c}{$R_0$}&\multicolumn{1}{c}{$(B-R)_0$}&&
\multicolumn{1}{c}{$R_0$}&\multicolumn{1}{c}{$(B-R)_0$}&&
\multicolumn{1}{c}{$R_0$}&\multicolumn{1}{c}{$(B-R)_0$}\\
\midrule
PRC A-01 &&$14.62 \pm 0.01$&$+1.62 \pm 0.02$ &&$14.44$&$+1.62$&&$16.23 \pm 0.03$&$+1.21 \pm 0.05$\\
PRC A-04 &&$14.72 \pm 0.01$&$+1.57 \pm 0.03$ &&$14.23$&$+1.53$&&$17.35 \pm 0.07$&$+1.10 \pm 0.15$\\
PRC A-06 &&$14.96 \pm 0.01$&$+1.59 \pm 0.02$ &&$14.66$&$+1.53$&&$16.58 \pm 0.03$&$+1.07 \pm 0.06$\\
PRC B-10 &&$16.14 \pm 0.02$&$+1.65 \pm 0.05$ &&$16.16$&$+1.58$&&$18.73 \pm 0.04$&$+1.49 \pm 0.16$\\
PRC B-17 &&$14.71 \pm 0.01$&$+0.89 \pm 0.03$ &&$13.69$&$+0.86$&&$16.97 \pm 0.04$&$+0.61 \pm 0.06$\\
PRC B-20 &&$14.68 \pm 0.01$&$+1.78 \pm 0.03$ &&$14.18$&$+1.67$&&$18.45 \pm 0.03$&$+1.67 \pm 0.10$\\
PRC B-21 &&$14.87 \pm 0.01$&$+1.77 \pm 0.03$ &&$14.22$&$+1.60$&&$17.00 \pm 0.02$&$+0.89 \pm 0.04$\\
\bottomrule
\tabnotetext{a}{Photometry in 15\farcs2 diameter circular aperture.}
\tabnotetext{b}{Photometry from Galfit.}
\tabnotetext{c}{Photometry in polygonal apertures over the ring. Note
  that these apertures do not
cover the whole ring and so the magnitudes given here underestimate the
total ring magnitude.}
\end{tabular}
\end{table*}

The host galaxies are not the focus of this work. Nevertheless, we
obtained photometry of the host galaxies primarily to permit a
comparison of our photometry with previous photometry (see \S4.4).

We obtained photometry for the host galaxies both from aperture
photometry and from the Galfit models. For the aperture photometry, we
first subtracted the sky as determined by Galfit and then used the
apphot package with apertures of diameter 15\farcs2, identical to those
used for our standard stars, with the background fixed at zero. For the
model photometry, we summed the components fitted by Galfit.
Table~\ref{table:photometry} shows the extinction-corrected $R_0$
magnitudes and $(B-R)_0$ colors for both methods. The model colors are
similar or slightly bluer than the aperture colors; this may reflect
color gradients in some of the galaxies.

We estimated uncertainties for our aperture photometry taking into
account the contribution of uncertainties in the calibration (2.0\% in
$B$ and 1.4\% in $R$), Poisson noise in the object and sky,
uncertainties in the flat field, and uncertainties in the determination
of the sky level. We estimated the later two by measuring the sky in
about a dozen small regions isolated from the galaxy and other sources
such as stars and calculating the standard deviation between these sky
measurements. In all galaxies except PRC B-10 in $B$, the dominant
contributor to the noise is the uncertainty in the calibration. In B-10
in $B$, the dominant contributor is uncertainty in the determination of
the sky level.

We have not calculated uncertainties for our photometry of the models
because the model photometry is peripheral to our aims here and because
obtaining uncertainties for the model parameters is involved.
Nevertheless, we expect them to be similar to the uncertainties in our
aperture photometry.

\subsection{Polar Rings}

\begin{table*}\centering
\setlength{\tabnotewidth}{\columnwidth}
\tablecols{2}
\caption{Offsets of the Vertices of the Ring Photometry Apertures}
\label{table:ver}
\begin{tabular}{ll}
\toprule
Galaxy & Vertices (West, North)\\
\midrule
PRC A-01 & $(-1\farcs0, +10\farcs0), (-9\farcs6, -1\farcs6), (-26\farcs8, +15\farcs6), (-22\farcs8, +22\farcs7)$\\
 & $(+10\farcs5, -0\farcs6), (+1\farcs4, -10\farcs2), (+19\farcs6, -22\farcs3), (+25\farcs7, -15\farcs3)$\\
\midrule
PRC A-04 & $(-8\farcs5, +10\farcs0), (-12\farcs5, +2\farcs9), (-37\farcs3, +18\farcs1), (-34\farcs3, +25\farcs2)$\\
 & $(+10\farcs6, -2\farcs9), (+6\farcs6, -10\farcs5), (+31\farcs4, -28\farcs2), (+34\farcs9, -21\farcs6)$  \\
\midrule
PRC A-06 & $(+3\farcs1, +17\farcs1), (-12\farcs1, +8\farcs0), (-18\farcs6, +39\farcs9), (-10\farcs0, +42\farcs4)$\\
 & $(-3\farcs3, -17\farcs6), (+10\farcs9, -10\farcs6), (+15\farcs5, -35\farcs9), (+5\farcs3, -40\farcs4), (+0\farcs8, -29\farcs8), (+3\farcs8, -23\farcs2)$  \\
\midrule
PRC B-10 &$(+1\farcs6, +5\farcs6), (-4\farcs4, +3\farcs6), (-4\farcs4, +11\farcs2), (-0\farcs4, +11\farcs7)$ \\
 & $(-1\farcs9, -6\farcs5), (+3\farcs1, -4\farcs5), (+3\farcs1, -10\farcs1), (+0\farcs1, -11\farcs6)$  \\
\midrule
PRC B-17 & $(-3\farcs7, +13\farcs4), (-13\farcs8, +7\farcs8), (-19\farcs4, +25\farcs0), (-8\farcs3, +25\farcs0)$ \\
 & $(+0\farcs6, -14\farcs9), (+11\farcs7, -10\farcs8), (+20\farcs3, -23\farcs5), (+13\farcs3, -27\farcs5)$  \\
\midrule
PRC B-20 &$(-8\farcs7, -2\farcs2), (-15\farcs8, -1\farcs7), (-15\farcs3, +3\farcs3), (-8\farcs2, +4\farcs8)$ \\
 & $(+6\farcs2, -6\farcs0), (+7\farcs2, +2\farcs1), (+14\farcs3, +0\farcs5), (+13\farcs3, -4\farcs5)$  \\
\midrule
PRC B-21 & $(-10\farcs8, -12\farcs2), (-22\farcs5, -13\farcs7), (-25\farcs0, -9\farcs2), (-15\farcs4, -4\farcs1)$ \\
 & $(+14\farcs2, -1\farcs6), (+9\farcs7, +8\farcs0), (+23\farcs3, +8\farcs0), (+24\farcs9, +0\farcs9)$  \\
\bottomrule
\end{tabular}
\end{table*}

To measure the ring colors, we obtained aperture photometry of the rings
over regions uncontaminated by the host galaxy. We used the images after
subtracting the sky and model determined by Galfit. We defined polygonal
apertures on the brightest and most isolated parts of the ring and used
the polyphot task in the apphot package with the background fixed at
zero. Table~\ref{table:ver} lists the offsets west and north from the
center of the galaxies of the vertices of the apertures.
Table~\ref{table:photometry} shows the ring aperture magnitudes and
colors corrected for Galactic extinction.

We estimated the uncertainties in the same way as for the aperture
photometry of the host galaxies. In all cases the dominant contributor
is uncertainty in the determination of the sky level.

Galfit can fit the sky using a tilted plane, which provides a clear
improvement over a simple constant. For example, if we had used a
constant rather than a tilted plane, our uncertainty in the $B-R$ color
of the ring of PRC B-20 would have been 0.34 rather than 0.10.
Furthermore, our use of polygonal apertures rather than circular
apertures also allowed us to reduce the uncertainty associated with the
sky level. Reshetnikov et al.\@ (1994) used a pair of 40{\arcsec}
diameter apertures for their photometry of the ring of PRC A-06. These
apertures have a combined area that is about 3.5 times larger than our
apertures. If we had used these apertures, our uncertainty in the $B-R$
color of the ring of PRC A-06 would have been 0.21 rather than 0.06.

\subsection{Approximate Color Transformations}

We are faced with the problem that different authors measure
different colors. To facilitate comparison, we have converted other
colors to $B-R$ using the approximate transformations
\begin{eqnarray}
(B-R) &\approx& 1.60 \pm 0.03 (B-V),\\
(B-R) &\approx& 1.29 \pm 0.07 (V-I),\\
(B-R) &\approx& 0.71 \pm 0.02 (B-I).
\end{eqnarray}
The coefficients are the mean coefficients for populations with ages of
$10^9$, $5 \times 10^9$, and $10^{10}$ years and metallicities $Z$ of
0.004, 0.008, 0.02, and 0.05 according to the models of Kurth et al.\
(1999). We also show the dispersion of the 12 individual coefficients
around these mean coefficients. These approximate transformations are
appropriate for populations dominated by intermediate-age and old stars.

\subsection{Comparison with Previous Galaxy Photometry}

\begin{table*}[p]
\centering
\setlength{\tabnotewidth}{0.7\linewidth}
\tablecols{4}
\caption{Comparison of Host Galaxy $(B-R)_0$ Colors}
\label{table:galaxy-colors-compared}
\begin{tabular}{lccc}
\toprule
Galaxy &
Our Aperture&
Our Model&
Others\\
\midrule
PRC A-01 &$+1.61 \pm 0.03$&$+1.61$&$+1.3$\tabnotemark{ab}\\
PRC A-02 &\nodata         &\nodata&$+1.6$\tabnotemark{ac},$+1.4$\tabnotemark{ac}\\
PRC A-03 &\nodata         &\nodata&$+1.4$\tabnotemark{ad}\\
PRC A-04 &$+1.57 \pm 0.03$&$+1.53$&$+1.75$\tabnotemark{e},$+1.77$\tabnotemark{f},$+1.30$\tabnotemark{g}\\
PRC A-05 &\nodata         &\nodata&$+1.3$\tabnotemark{ah},$+1.4$\tabnotemark{ai}\\
PRC A-06 &$+1.58 \pm 0.03$&$+1.52$&$+1.47$\tabnotemark{j}\\
PRC B-03 &\nodata         &\nodata&$+1.39$\tabnotemark{k}\\
PRC B-10 &$+1.63 \pm 0.05$&$+1.56$&\nodata\\
PRC B-17 &$+0.88 \pm 0.03$&$+0.85$&$+0.99$\tabnotemark{l}, $+0.9\pm0.2$\tabnotemark{m}\\
PRC B-19 &\nodata         &\nodata&$+1.6$\tabnotemark{n}\\
PRC B-20 &$+1.76 \pm 0.03$&$+1.65$&\nodata\\
PRC B-21 &$+1.76 \pm 0.03$&$+1.59$&$+1.4$\tabnotemark{o},$+1.5 \pm 0.2$\tabnotemark{p}, $0.90\pm0.1$\tabnotemark{q}\\
PRC C-02 &\nodata         &\nodata&$+0.90$\tabnotemark{r}\\
PRC C-03 &\nodata	  &\nodata&$+1.31$\tabnotemark{s}\\
PRC C-12 &\nodata         &\nodata&$+0.95$\tabnotemark{r}\\
PRC C-13 &\nodata         &\nodata&$+2.3$\tabnotemark{at}\\
PRC C-27 &\nodata         &\nodata&$+0.73$\tabnotemark{r}\\
\bottomrule
\tabnotetext{a}{Color converted to $B-R$ using
  the approximate transformations of \S4.4. The details of the
  conversions are mentioned in the notes below.}
\tabnotetext{b}{Reshetnikov et al. (1994), converted from $B-V = +0.8$
  obtained in turn by these authors by converting the $B-g$ and $B-i$ colors of Whitmore
et al. (1990).}
\tabnotetext{c}{Whitmore et al. (1987), converted from $B-V = 1.02 \pm 0.03$ in an aperture of 2\farcs9
  over the nucleus and $B-V = 0.90 \pm 0.08$ in the halo.}
\tabnotetext{d}{Peletier \& Christodoulou
  (1993), converted from $B-V = 0.87$ outside the central 4\farcs0.}
\tabnotetext{e}{Mould et al. (1982), in an aperture of 8\farcs4 diameter.}
\tabnotetext{f}{Reshetnikov et al. (1994), in an aperture of 8\farcs4 diameter.}
\tabnotetext{g}{Reshetnikov et al. (1994), total magnitude.}
\tabnotetext{h}{Whitmore et al. (1987), converted from $B-V = +0.81$
  obtained in turn by these authors from Sérsic \& Agüero (1972).}
\tabnotetext{i}{Gallagher et al. (2002), converted from $B-I=2.0$.}
\tabnotetext{j}{Reshetnikov et al. (1994).}
\tabnotetext{k}{Reshetnikov et al. (1995).}
\tabnotetext{l}{Cox et al. (2001).}
\tabnotetext{m}{Gil de Paz et al. (2003).}
\tabnotetext{n}{Peletier et al. (1990).}
\tabnotetext{o}{Stockton \& MacKenty (1983).}
\tabnotetext{p}{Lauberts \& Valentijn (1989).}
\tabnotetext{q}{Reshetnikov et al. (2002).}
\tabnotetext{r}{Reshetnikov (2004).}
\tabnotetext{s}{Reshetnikov et al. (2005), using an aperture of
  20\farcs0 diameter.}
\tabnotetext{t}{van Driel et al. (1995), converted from $V-I = +1.8$.}
\end{tabular}
\end{table*}

Table~\ref{table:galaxy-colors-compared} compares our photometry for the
host galaxies with that of previous authors. Comparing galaxy colors
directly is made difficult by the fact that different workers have used
different measurement apertures. Because galaxies often show intrinsic
radial color gradients, this can lead to differences in the measured
aperture colors at the tenth of a magnitude level. However, we can make
a direct comparison of our photometry with that of Mould et al.\@ (1982)
and Reshetnikov et al.\@ (1994) for PRC A-04. These authors give $B-R$
colors of +1.75 and +1.77 in a 8\farcs4 diameter aperture. We measure a
color of $+1.66 \pm 0.03$ in this aperture. Unfortunately, the other
authors do not give any indications of the likely uncertainties in their
measurements, and this makes it impossible to evaluate the difference of
about 0.1 magnitudes between our measurement and their measurements.
However, if their measurements have similar errors to our measurement,
then the difference is only about $2\sigma$ and as such is not
significant.

Our color for B-17 is consistent with the somewhat uncertain color given
by Gil de Paz et al.\@ (2003). Our color is slightly bluer than that
+0.99 (corrected for Galactic extinction) given by Cox et al.\@ (2001).
However, their color was measured on patches of the galaxy away from the
ring, whereas our color includes contamination from the bluer ring.

The two worst differences are A-01 and B-21. In the case of A-01 our
$B-R$ color is 0.3 magnitudes redder than would be expected from the $B$
and $g$ photometry of Whitmore et al.\@ (1990) converted to $B-V$ by
Reshetnikov et al.\@ (1994) and finally to $B-R$ using the approximate
transformations above. Interestingly, we both measure the ring to be
about 0.4 magnitudes bluer than the host galaxy, which might suggest a
problem with the photometic calibration. Unfortunately, Whitmore et
al.\@ (1990) give no details of their photometric calibration, instead
refering the reader Whitmore et al.\@ (1987). In that paper, the authors
use mean extinction coefficients, although it seems unlikely that this
could explain all of the difference.

In the case of B-21, our model $B-R$ color is fully 0.7 magnitudes
redder than the $B-R$ color measured by Reshetnikov et al.\@ (2002), but
is in good agreement with the color given by Lauberts \& Valentijn
(1989) and reasonable agreement with the color given by Stockton \&
MacKenty (1983). Reshetnikov et al.\@ (2002) measure similar colors for
the host galaxy and the ring, but we measure much bluer colors for the
ring. We conclude that the photometry of Reshetnikov et al.\@ (2002) is
probably in error. We note again these authors used mean extinction
coefficients, although again this is unlikely to lead to an error as
large as 0.7 magnitudes.

We conclude that our colors are consistent with previous authors to
around one tenth of a magnitude or better, with the exception of the
colors of A-01 and B-21 measured by Whitmore et al.\@ (1990) and
Reshetnikov et al.\@ (2002). We believe that the internal dispersion in
our photometric standards of 0.024 magnitudes in $B-R$ is a reliable
reflection of the accuracy of our photometic calibration.

\subsection{Comparison with Previous Ring Colors}

\begin{table}\centering
\setlength{\tabnotewidth}{\columnwidth}
\tablecols{3}
\caption{Comparison of Ring Colors}
\label{table:ring-colors-compared}
\begin{tabular}{lcc}
\toprule
Galaxy &
Our $(B-R)_0$&
Others $(B-R)_0$\\
\midrule
PRC A-01 &$+1.20 \pm 0.05$&$+0.9$\tabnotemark{ab}\\
PRC A-02 &\nodata         &$+1.0$\tabnotemark{ac}\\
PRC A-03 &\nodata         &$+1.1$\tabnotemark{ad}\\
PRC A-04 &$+1.10 \pm 0.15$&$+0.82$\tabnotemark{e}, $+1.0\pm0.1$\tabnotemark{f}\\
PRC A-05 &\nodata         &$-0.1$\tabnotemark{ag},$+1.1$\tabnotemark{ah}\\
PRC A-06 &$+1.07 \pm 0.06$&$+0.91$\tabnotemark{i}\\
PRC B-03 &\nodata         &$+0.6\pm0.14$\tabnotemark{j}\\
PRC B-10 &$+1.47 \pm 0.16$&\nodata\\
PRC B-17 &$+0.61 \pm 0.06$&$+0.63$\tabnotemark{k}\\
PRC B-19 &\nodata         &$+1.0$\tabnotemark{l}\\
PRC B-20 &$+1.65 \pm 0.10$&\nodata\\
PRC B-21 &$+0.88 \pm 0.04$&$+0.5$\tabnotemark{m}\\
PRC C-02 &\nodata         &$+0.86$\tabnotemark{n}\\
PRC C-03 &\nodata	  &$+0.9$\tabnotemark{o}\\
PRC C-13 &\nodata         &$+1.3$\tabnotemark{ap}\\
PRC C-27 &\nodata         &$+0.6$\tabnotemark{n}\\
\bottomrule
\tabnotetext{a}{Color converted to $B-R$ using
  the approximate transformations of \S4.4. The details of the
  conversions are mentioned in the notes below.}
\tabnotetext{b}{Reshetnikov et al. (1994), converted from $B-V = +0.55$
  obtained in turn by these authors from the $B-g$ and $B-i$ of Whitmore
et al. (1990).}
\tabnotetext{c}{Whitmore et al. (1987), converted from $B-V = +0.65 \pm 0.08$.}
\tabnotetext{d}{Peletier \& Christodoulou (1993), converted from  $B-V =
  0.71$.}
\tabnotetext{e}{Reshetnikov et al. (1994).}
\tabnotetext{f}{Mould et al. (1982).}
\tabnotetext{g}{Whitmore et al. (1987), converted from $B-V = -0.09$
  obtained in turn by these authors from Sérsic \& Agüero (1972).}
\tabnotetext{h}{Gallagher et al. (2002), converted from $B-I \approx 1.5$.}
\tabnotetext{i}{Reshetnikov et al. (1994).}
\tabnotetext{j}{Reshetnikov et al. (1995).}
\tabnotetext{k}{Cox et al. (2001).}
\tabnotetext{l}{Arnaboldi et al. (1993).}
\tabnotetext{m}{Reshetnikov et al. (2002).}
\tabnotetext{n}{Reshetnikov (2004).}
\tabnotetext{o}{Reshetnikov et al. (2005).}
\tabnotetext{p}{van Driel et al. (1995), converted from $V-I = 1.0$.}
\end{tabular}
\end{table}

Table~\ref{table:ring-colors-compared} compares our photometry for the
rings with that of previous authors. We are in good agreement with
previous authors on the colors of PRC A-04 and B-17 and reasonable
agreement with previous authors on the colors of A-06. We note that for
their photometry of A-06, Reshetnikov et al.\@ (1994) used much larger
apertures the we have used, and as such their colors may be more
sensitive to uncertainties in the determination of the sky level. Our
colors for A-01 and B-21 are significantly redder than the colors given
by Reshetnikov et al.\@ (1994) and Reshetnikov et al.\@ (2002), but
since our photometry of the host galaxy is also different, this is no
surprise.

\section{Discussion}

\subsection{Ring Magnitudes}

Our ring magnitudes in $R$ are 1.8 to 4.3 magnitudes fainter than our
model magnitudes for the host galaxies. Of course, our ring magnitudes
do not include the whole ring. We can make a rough correction to this by
assuming that they include half of the ring, in which case the rings are
1.1 to 3.6 magnitudes fainter than the galaxies. Thus, we estimate that
these rings contribute between 4\% and 40\% of the light of the host
galaxy, which is in agreement with other authors (approximately 25\% in
PRC A-05, Gallagher et al. 2002; 11\% in PRC C-02, Reshetnikov 2004;
14\% for PRC A-04 and 24\% for PRC A-06, Reshetnikov \& Combes 1994;
between 1\% and 40\% for the confirmed polar-ring galaxies in the PRC,
Reshetnikov et al. 1994). The most dominant rings are in PRC A-01 and
A-06.

\subsection{Range of Ring Colors}

We have measured the $B-R$ colors of seven polar rings. Previous
observers have generally obtained ring colors bluer than $B-R = +1.1$,
with the notable exception of van Driel et al.\@ (1995) who obtained $V
- I \approx +1.0$ (which suggest $B-R \approx +1.5$) for the ring of PRC
C-13 (NGC 660).

Our measured $B-R$ ring colors span a range from $+0.61\pm0.06$ to
$+1.65\pm0.10$. The two reddest rings are PRC B-10 and B-20. In B-10,
the ring is faint and the color of $+1.47\pm0.16$ is correspondingly
uncertain. The color of the ring of B-10 is consistent at the $3\sigma$
level with an intermediate color $+1.0$. On the other hand, in B-20 the
ring is brighter and has better measured color of $+1.65\pm0.10$, which
even with a $3\sigma$ departure to the blue would still be as red as
$+1.35$. Thus, we consider that B-20 has a truly red ring.

Thus, our work identifies B-20 as a second galaxy with a red ring and
confirms that rings have a spread in colors. This spread presumably is
caused by differences in one or more of the ring ages, metallicities, and
internal reddenings.

\subsection{The Blue Rings of PRC B-17 and B-21}

Our bluest two rings, those of PRC B-17 and B-21, are the only ones in
our sample that have very bright \ion{H}{2} regions (Cox et al.\@ 2001;
Watson, unpublished). This suggests that recent star formation is the
cause of these blue colors.

\subsection{The Red Ring of PRC B-20}

As we discussed above, the ring of PRC B-20 has a $B-R$ color of
$+1.65\pm0.10$. A stellar population may show red colors primarily
because it is old or because it suffers reddening. Metallicity also
modifies the colors, especially for older populations.

Unfortunately, the internal extinction of polar rings is not well
understood. Gallagher et al.\ (2002) formed a color-magnitude diagram
for the ring of PRC A-05 on a pixel-to-pixel basis. Dusty pixels were
statistically redder and fainter. Unfortunately our data do not have
adequate spatial resolution to apply this technique. Furthermore, the
technique is insensitive to a uniform extinction.

Additionally, the metallicity of the ring of B-20 is unknown. However,
we note that there is distinct absorption from the ring where it passes
over the south side of the galaxy. Therefore, it seems unlikely that the
ring is especially metal poor. In other rings, the metallicity ranges
from about solar (Eskridge \& Pogge 1997) to about half solar
(Buttiglione, Arnaboldi, \& Iodice 2006).

Finally, the transformation between the current colors of a stellar
population and its age depend on the detailed star formation history.
For example, the ring could have existed for a certain time before it
began to form stars. Alternatively, it may have formed stars
continuously rather than in a single burst, in which case its colors
would be bluer. In our analysis here, we assume that the stars formed in
a single burst, which gives a minimum ring age.

We can estimate possible minimum ages for the ring under several
different assumptions. We assume the metallicity can range from
two-fifths solar ($Z=0.008$) to solar ($Z=0.02$) and the internal
reddening from 0 to 0.3 magnitudes. An intrinsic color of $+1.65$ would
imply a minimum age in excess of 10 Gyr according to the models of Kurth
et al.\@ (1999), irrespective of metallicity. On the other hand, an
intrinsic color of $+1.35$, arising either from a $3\sigma$ departure to
the blue or $+0.3$ magnitudes of internal extinction, would imply
minimum ages of about 2.5 Gyr for solar metallicity and 5 Gyr for
two-fifths solar metallicity. Finally, an intrinsic color of $+1.05$,
arising from both $3\sigma$ departure to the blue and $+0.3$ magnitudes
of internal extinction, would imply minimum ages of about 1.2 Gyr for
solar metallicity and 1.7 Gyr for two-fifths solar metallicity.


With a distance of about 140 Mpc (Jones et al.\@ 2004, assuming $H_0 =
75\,\mathrm{km\,s^{-1}\,Mpc,s^{-1}}$), the ring diameter of B-20 of
about 22{\arcsec} corresponds to about 15 kpc. If the orbital velocity
at this distance from the galaxy is $120\,\mathrm{km\,s^{-1}}$, the
orbital period is about 0.4 Gyr. If the ring in B-20 is definitely older
than 1.2 Gyr, the ring has survived for at least 3 orbital periods and
so would appear to be stable.

On the other hand, before we rush to conclude that all polar rings are
stable, we must note that the ring in B-20 has not been confirmed as a
polar ring; it is simply a good candidate (and, for that matter, neither
has the ring in B-10). Confirmation will require measurements of the
kinematics of the host galaxy and the ring. Furthermore, our minimum age
estimate is severely compromised by uncertainties in the metallicity and
internal reddening of the ring. However, future spectroscopy could yield
a ring metallicity and imaging in the near-infrared might reduce the
uncertainty due to internal reddening. These observations should be
pursued.

\acknowledgements

We thank Pedro Colín, José Antonio de Diego, Simon Kemp, and Michael
Richer for useful comments on an early version of this work. We thank an
anonymous referee for a useful report. We are grateful to the staff of
the Observatorio Astronómico Nacional on Sierra San Pedro Mártir for
their hospitality and technical support during the observing runs. This
work has been supported in part by the project IN118302-3 of the
UNAM/DGAPA. This research has made use of the NASA/IPAC Extragalactic
Database (NED) which is operated by the Jet Propulsion Laboratory,
California Institute of Technology, under contract with the National
Aeronautics and Space Administration.



\newcommand{\images}[5]{
\begin{figure*}[p]\centering

\scalebox{-1}[1]{\includegraphics[angle=-90,width=0.24\textwidth]{#1-b-1.jpg}}
\scalebox{-1}[1]{\includegraphics[angle=-90,width=0.24\textwidth]{#1-b-2.jpg}}
\scalebox{-1}[1]{\includegraphics[angle=-90,width=0.24\textwidth]{#1-b-3.jpg}}
\scalebox{-1}[1]{\includegraphics[angle=-90,width=0.24\textwidth]{#1-b-4.jpg}}

\scalebox{-1}[1]{\includegraphics[angle=-90,width=0.24\textwidth]{#1-r-1.jpg}}
\scalebox{-1}[1]{\includegraphics[angle=-90,width=0.24\textwidth]{#1-r-2.jpg}}
\scalebox{-1}[1]{\includegraphics[angle=-90,width=0.24\textwidth]{#1-r-3.jpg}}
\scalebox{-1}[1]{\includegraphics[angle=-90,width=0.24\textwidth]{#1-r-4.jpg}}

\caption{Images of PRC #2 in $B2$ (top row) and $R2$ (bottom row). The images
  are $#3 \arcsec \times #4 \arcsec$. #5}
\label{figure:#2}

\end{figure*}
}

\clearpage

\images{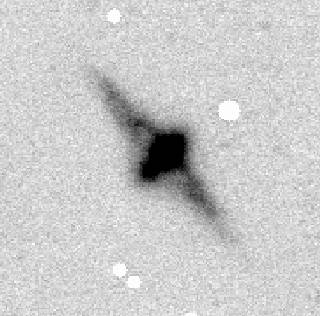}{A-01}{90}{90}{North is up and east is left. The
  first column shows the host galaxy and the ring. The second column
  shows the mask used to discard certain regions from the model fit to
  the host galaxy. The third column shows the ring after the model of
  the host galaxy is subtracted. The fourth column shows the apertures
  used for photometry of the ring.}

\images{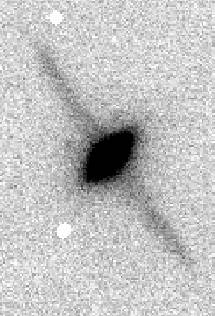}{A-04}{90}{60}{Otherwise as Figure 1.}
\images{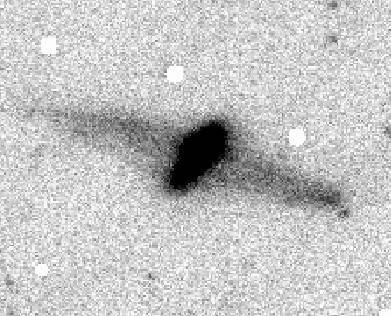}{A-06}{90}{110}{Otherwise as Figure 1.}
\images{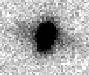}{B-10}{21}{25}{Otherwise as Figure 1.}
\images{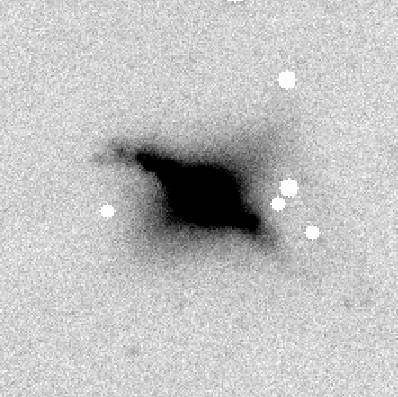}{B-17}{110}{110}{Otherwise as Figure 1.}
\images{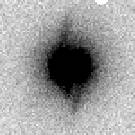}{B-20}{37}{37}{Otherwise as Figure 1.}
\images{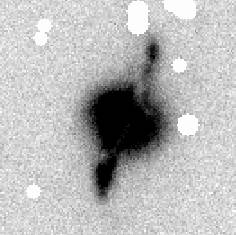}{B-21}{65}{65}{Otherwise as Figure 1.}


\begin{thebibliography}

\bibitem{} Arnaboldi,~M., Capaccioli,~M., Cappellaro,~E., Held,~E.~V.,
Sparke,~L. 1993, A\&A, 267, 21

\bibitem{} Bournaud, F., \& Combes, F. 2003, A\&A, 401, 817

\bibitem{} Buttiglione, S., Arnaboldi, M., \& Iodice, E. 2006, Memorie
della Societa Astronomica Italiana Supplement, 9, 317

\bibitem{} Cox,~A.~L. \& Sparke,~L.~S. 1996, ASPC, 106, 168

\bibitem{} Cox,~A.~L., Sparke,~L.~S., Watson,~A.~M., \& van~Moorsel,~G.
2001, AJ, 121, 692

\bibitem{} Eskridge,~P.~B., \& Pogge,~R.~W. 1997, ApJ, 486, 259

\bibitem{} Gallagher,~J.~S., Sparke,~L.~S., Matthews,~L.~D.,
Frattare,~L.~M., English,~J., Kinney,~A.~L., Iodice,~E., Arnaboldi,~M.
2002, ApJ, 568, 199

\bibitem{} Gil~de~Paz,~A., Madore,~B.~F., \& Pevunova,~O. 2003, ApJS,
147, 29

\bibitem{} Jones,~D.~H., Saunders,~W., Colless,~M., Read,~M.~A.,
Parker,~Q.~A., Watson,~F.~G., Campbell,~L.~A., Burkey,~D., Mauch,~T.,
Moore,~L.~, Hartley,~M.~, Cass,~P., James,~D.~, Russell,~K.,
Fiegert,~K., Dawe,~J., Huchra,~J., Jarrett,~T., Lahav,~O., Lucey,~J.,
Mamon,~G.~A., Proust,~D., Sadler,~E.~M., \& Wakamatsu,~K. 2004, MNRAS,
355, 747

\bibitem{} Kurth,~O.~M., Fritze-v.~Alvensleben,~U., \& Fricke,~K.~J.
1999, A\&AS, 138, 19

\bibitem{} Landolt,~A.~U. 1983, AJ, 88, 439L

\bibitem{} Lauberts, A. \& Valentijn, E. A. 1989, The Surface Photometry
Catalogue of the ESO-Uppsala Galaxies

\bibitem{} Mould,~J., Balick,~B. \& Aaronson,~M. 1982, ApJ 260, L37

\bibitem{} Peletier~R., Davies~R.~L., Illingworth~G.~D., Davis~L.~E.,
Cawson~M. 1990, AJ 100, 1091

\bibitem{} Peletier,~R.~F., Christodoulou,~D.~M. 1993, AJ, 105, 1378

\bibitem{} Peng,~C.~Y., Ho,~L.~C., Impey,~C.~D. \& Rix,~H.~W. 2002, AJ,
124, 266

\bibitem{} Reshetnikov,~V.~P. \& Combes,~F. 1994 A\&A, 291, 57

\bibitem{} Reshetnikov,~V.~P., Hagen-Thorn,~V.~A., \& Yakovleva,~V.~A.
1994, A\&A, 290, 693

\bibitem{} Reshetnikov,~V.~P., Hagen-Thorn,~V.~A., \& Yakovleva,~V.~A.
1995, A\&A, 303, 398

\bibitem{} Reshetnikov,~V.~P., Faúndez-Abans,~M., \&
de~Oliveira-Abans,~M. 2002, A\&A, 383, 390

\bibitem{} Reshetnikov,~V.~P. 2004, A\&A, 416, 889

\bibitem{} Reshetnikov,~V., Bournaud,~F., Combes,~F., Faúndez-Abans,~M.,
de~Oliveira-Abans,~M., van~Driel,~W., \& Schneider,~S.~E. 2005, A\&A,
431, 503

\bibitem{} Schechter,~P.~L. \& Gunn,~J.~L. 1978, AJ, 83, 1360

\bibitem{} Schlegel,~D.~J., Finkbeiner,~D.~P., Davis,~M. 1998, ApJ, 500,
525

\bibitem{} Schweizer, F., Whitmore, B.~C., \& Rubin, V.~C. 1983, AJ, 88, 909

\bibitem{} Sérsic,~J.~L. \& Agüero,~E.~L. 1972, Ap\&SS, 19, 387

\bibitem{} Sparke,~L.~S. 2004, ``WARPS, Polar Rings, and High-Velocity
  Clouds'', eds.\ H. van Woerden, B. P. Wakker, U. J. Schwarz, \& K. S.
  de Boer (Dordrech: Kluwer Academic Publishers), Astrophysics and Space
  Science Library 213, 273.

\bibitem{} Stockton, A. \& MacKenty, J. W. 1983, Nature 305, 678


\bibitem{} van~Driel,~W., Combes,~F., Casoli,~F., Gerin,~M., Nakai,~N.,
Miyaji,~T., Hamabe,~M., Sofue,~Y., Ichikawa,~T., Yoshida,~S.,
Kobayashi,~Y., Geng,~F., Minezaki,~T., Arimoto,~N., Kodama,~T.,
Goudfrooij,~P., Mulder,~P.~S., Wakamatsu,~K., \& Yanagisawa,~K. 1995, AJ,
109, 942

\bibitem{} Whitmore,~B.~C., Lucas,~R.~A., McElroy,~D.~B.,
Steiman-Cameron,~T.~Y., Sackett,~P.~D. \& Olling,~R.~P. 1990, AJ, 100,
1489

\bibitem{} Whitmore,~B.~C., McElroy,~D.~B. \& Schweizer,~F. 1987, AJ,
314, 439






%
%
%
%
%
%

\end{thebibliography}
\end{document}